\documentclass[runningheads]{llncs}
\usepackage[T1]{fontenc}
\usepackage{amsmath}
\usepackage{graphicx}
\usepackage{amsfonts} 
\usepackage{amsmath}
\usepackage[table,xcdraw]{xcolor}
\usepackage{array}
\usepackage{caption}
\usepackage{floatrow}   
\usepackage{subcaption}
\usepackage{graphicx,xcolor} 
\usepackage{breqn}
\usepackage[framemethod=tikz]{mdframed}
\usepackage[tableposition=top]{caption}
\usepackage{float}
\floatstyle{plaintop}
\restylefloat{table}
\usepackage{arydshln}
\usepackage{hyperref}
\usepackage{color, colortbl}
\definecolor{Gray}{gray}{0.85}
\usepackage{booktabs}
\usepackage[normalem]{ulem}
\usepackage{graphicx}
\usepackage{subcaption}
\begin{document}
	\title{Physically Inspired Constraint for Unsupervised Regularized Ultrasound Elastography\thanks{Supported by Natural Sciences and Engineering Research Council of Canada (NSERC) Discovery Grant. The Alpinion ultrasound machine was partly funded by Dr. Louis G. Johnson Foundation.}}
	%
	%
	\author{Ali K. Z. Tehrani$^{1,2}$, and Hassan Rivaz$^{1,3}$}
	\authorrunning{Ali K. Z. Tehrani}
	%
	\institute{Department of Electrical and Computer Engineering, Concordia University, Canada\and 
		\email{a_kafaei@encs.concordia.ca}  \email{ $^3$ hrivaz@ece.concordia.ca}}
	\maketitle              
	\begin{abstract}
		Displacement estimation is a critical step of virtually all Ultrasound Elastography (USE) techniques. Two main features make this task unique compared to the general optical flow problem: the high-frequency nature of ultrasound radio-frequency (RF) data and the governing laws of physics on the displacement field. Recently, the architecture of the optical flow networks has been modified to be able to use RF data. Also, semi-supervised and unsupervised techniques have been employed for USE by considering prior knowledge of displacement continuity in the form of the first- and second-derivative regularizers. Despite these attempts, no work has considered the tissue compression pattern, and displacements in axial and lateral directions have been assumed to be independent. However, tissue motion pattern is governed by laws of physics in USE, rendering the axial and the lateral displacements highly correlated. In this paper, we propose Physically Inspired ConsTraint for Unsupervised Regularized Elastography (PICTURE), where we impose constraints on the Poisson’s ratio to improve lateral displacement estimates. Experiments on phantom and \textit{in vivo} data show that PICTURE substantially improves the quality of the lateral displacement estimation.
		
		\keywords{Ultrasound elastography  \and Physically inspired learning \and  Unsupervised training  }
	\end{abstract}
	\section{Introduction}
	
	Ultrasound (US) imaging is a popular modality due to its portability and ease-of-use especially in image-guided interventions. Ultrasound Elastography (USE) aims to provide stiffness information of the tissue. In USE, an external or internal force deforms the tissue, and US images before and after the tissue deformation are compared to obtain the displacement map. The displacement map is employed to obtain the strain map which reveals the elastic properties of the tissue. Free-hand palpation is a common USE method wherein the external force is applied by the operator using the probe without the need of any external hardware. The quality of the obtained strain map heavily relies on the accuracy of the estimated displacement and many different methods have been developed to obtain high-quality displacement estimation \cite{JHall2011,mirzaei2019combining}.  
	
	Convolutional Neural Networks (CNN) have shown promising results in optical flow problems. They have been successfully adopted for USE by modifying the architectures to handle high-frequency radio-frequency (RF) data \cite{tehrani2020displacement,tehrani2021mpwc}. Unsupervised and semi-supervised techniques have also been employed to train the networks using real US data without requiring the ground truth displacements \cite{tehrani2020semi,delaunay2020unsupervised,delaunay2021unsupervised,tehrani2022bi}. They employed prior knowledge of displacement continuity in the form of the first- and second-derivative regularizers. Despite these attempts, no work has ever considered the tissue compression pattern. The assumptions on the motions in USE can be utilized to provide prior information about the lateral displacement, which is usually poor in free-hand palpation compared to the axial one. In this paper, we propose Physically Inspired ConsTraint for Unsupervised Regularized Elastography (PICTURE), inspired by Hooke’s law and constraints on the Poisson’s ratio to improve the lateral displacement using the prior knowledge of the compression physics. We show that PICTURE substantially improves the lateral displacement estimation using constraints on the Poisson’s ratio.    
	\section{Method}
	\subsection{Hook's Law and Poisson's Ratio}
	\subsubsection{Homogeneous Material:}
	Assuming linear elastic, isotropic, and homogeneous material, Hooke's law can be written for 3 dimensions as \cite{ugural2003advanced}:  
	
	\begin{equation}
		\label{eq:hook}
		\begin{bmatrix}
			\varepsilon_{11}\\ 
			\varepsilon_{22}\\ 
			\varepsilon_{33}\\ 
			2\varepsilon_{23}\\ 
			2\varepsilon_{13}\\ 
			2\varepsilon_{12}
		\end{bmatrix}
		= \frac{1}{E}
		\begin{bmatrix}
			1 & -v & -v & 0 & 0 & 0\\ 
			-v& 1 & -v & 0 & 0 & 0\\ 
			-v&  -v&  1&  0&  0& 0\\ 
			0&  0&  0&  2+2v&  0& 0\\ 
			0&  0&  0&  0&  2+2v& 0\\ 
			0&  0&  0&  0&  0&2+2v 
		\end{bmatrix}
		\begin{bmatrix}
			\sigma_{11}\\ 
			\sigma_{22}\\ 
			\sigma_{33}\\ 
			\sigma_{23}\\ 
			\sigma_{13}\\ 
			\sigma_{12}
		\end{bmatrix}
	\end{equation}
	
	\noindent where $\varepsilon$, $E$ and $\sigma$ represent strain, Young’s modulus, and stress, respectively. Also, $v$ is the Poisson's ratio, and depends on the material. $\varepsilon_{ij}$ can be obtained by taking the derivative of the displacement in direction $i$ ($W_i$) with respect to the direction $j$
	\begin{equation}
		\varepsilon_{ij} = \frac{\partial W_i}{\partial j}
	\end{equation}
	where subscripts 1, 2, and 3 denote axial, lateral, and out-of-plane directions, respectively. 
	When a material is compressed in one direction (in USE, it is often the axial direction), it expands in the other directions (in USE, they are lateral and out-of-plane directions). Considering this effect, the lateral and axial strains are not independent anymore. The uniaxial stress can be assumed in free-hand palpation since the operator compresses the tissue downwards and shear stresses on the top are negligible \cite{ugural2003advanced}. This assumption leads to the simplification of Eq \ref{eq:hook}. All stress components except $\sigma_{11}$ can be ignored. The strain components are obtained by:
	\begin{equation}      
		\varepsilon_{11} = \frac{\sigma_{11}}{E}, \varepsilon_{22} = -v\frac{\sigma_{11}}{E},  \varepsilon_{33} = -v\frac{\sigma_{11}}{E} 
	\end{equation}
	This equation indicates that the lateral strain ($\varepsilon_{22}$) can be directly obtained from the axial strain ($\varepsilon_{11}$) and the Poisson's ratio ($\varepsilon_{22} = -v\times\varepsilon_{11}$).
	\subsubsection{Inhomogeneous Material:} When the material is inhomogeneous, the lateral strain ($\varepsilon_{22}$) cannot be directly obtained from Poisson's ratio and the axial strain ($\varepsilon_{11}$) since the assumption of uniaxial compression does not hold everywhere, especially on the borders of the inclusions. Total strain ($\varepsilon_{ij}$) in this case, is obtained by adding the contribution of elastic strain ($e_{ij}$) and non-elastic eigenstrain ($\varepsilon_{ij}^{\ast}$) \cite{ma2014principle}:
	\begin{equation}
		\varepsilon_{ij} = 	e_{ij} + \varepsilon_{ij}^{\ast}
	\end{equation}
	Eigenstrain is introduced to model the variation of total strain from elastic strain in the presence of inhomogeneity. It is zero inside the inclusions and decays toward zero with increasing distance from the inclusion boundaries \cite{ma2014principle}.
	Despite not being able to directly estimate $\varepsilon_{22}$ from $\varepsilon_{11}$ due to having inhomogeneity (eigenstrain),  they are highly correlated and $\varepsilon_{11}$ can provide prior information about $\varepsilon_{22}$. Effective Poisson Ratio (EPR) is defined as \cite{islam2018new}:
	\begin{equation}
		\label{eq:epr}
		v_e = \frac{-\varepsilon_{22}}{\varepsilon_{11}}
	\end{equation}
	where EPR ($v_e$) is obtained by point-by-point division of the lateral and axial strains.  
	Although in inhomogeneous tissues, the EPR differs from the Poisson's ratio of the tissue and temporally, spatially variant, it is still a suitable approximator of the Poisson's ratio. It has been used to characterize tissues \cite{islam2018new}. The Poisson's ratio range under an arbitrary type of deformation and loading is between 0.2 and 0.5 \cite{mott2013limits}. Also, EPR has the same range as Poisson's ratio \cite{righetti2004feasibility}. Therefore, this range can be utilized to improve the lateral displacement quality.
	\subsection{Physically Inspired ConsTraint for Unsupervised Regularized Elastography (PICTURE)}
	We propose to utilize EPR as a prior information to improve the estimation of the lateral displacement, which is usually poor in USE. Although the ground-truth value of EPR is not known, the accepted range can be employed as a constraint in unsupervised training. We define a mask ($M$) to determine the EPR values outside the accepted range as:
	\begin{equation}
		M(i,j) = \begin{Bmatrix}
			0 &  & v_{emin}<\widetilde{v_e}(i,j)<v_{emax} \\
			1&  &  otherwise\\
		\end{Bmatrix}
	\end{equation}
	where $v_{emin}$ and $v_{emax}$ are the minimum and maximum accepted EPR values, and we assume the range to be between 0.1 to 0.6 (to have a small margin of error). In the next step, the following loss is employed to penalize the EPR values outside the accepted range as the following:
	\begin{equation}
		\begin{gathered}
			L_{vd} = \left| M\otimes (\varepsilon_{22}+<\widetilde{v_e}>\times \varepsilon_{11})\right|_2 
		\end{gathered}
	\end{equation}
	where $|.|_2$ denotes the L2 norm, $\otimes$ is the Kronecker product to select the incorrect EPR values, and $<\widetilde{v_e}>$ represents the average of EPR values in the accepted range (where $M=0$) which is obtained by:  
	\begin{equation}
		<\widetilde{v_e}> = \frac{\sum_{i,j}^{}(1-M_{(i,j)})\otimes v_e}{\sum_{i,j}^{}(1-M_{(i,j)})}
	\end{equation}
	The first-order derivative of $v_e$ are also added to enforce the smoothness of the EPR. 
	\begin{equation}
		L_{vs} = |\frac{\partial v_e}{\partial a}|_1 + \beta \times |\frac{\partial v_e}{\partial l}|_1
	\end{equation} 
	where $a$ and $l$ represent the axial and lateral directions, respectively. The parameter $\beta$ depends on the ratio of the spatial distance between two samples in axial and lateral directions, and is set it to 0.1 similar to \cite{tehrani2022bi}. 
	Finally, PICTURE loss is obtained as:
	\begin{equation} 
		L_V = L_{vd} + \lambda_{vs} \times L_{vs}
	\end{equation} 
	where $\lambda_{vs}$ is a hyper-parameter that controls the smoothness constraint.
	\subsection{Unsupervised training}
	Let $I_1,I_2\in \mathbb{R}^{3\times w \times h}$ be the pre-compression, and post-compression US data each has the width $h$, height $w$ and the 3 channels of RF data, the imaginary part of the analytic signal and the envelope of RF data, respectively (similar to \cite{tehrani2022bi}). The data loss in unsupervised training is the photometric loss obtained by comparing $I_1$ and warped $I_2$ ($\tilde{I_2}$) using bi-linear warping by the displacement $W$ which can be defined as \cite{tehrani2022bi,tehrani2020semi}:
	\begin{equation}
		\label{eq:loss_d}
		L_D =  |(I_1-\tilde{I_2})|_{1_{N\times N}} 
	\end{equation}
	where $|.|_1$ denotes the L1 norm, and a window of size $N\times N$ is considered around each sample to reduce the noise caused by warping. We simply set $N=3$ in this work. The data loss alone results in noisy displacement, and because the derivatives of the displacements are required in USE, smooth displacements are desired. Therefore, regularization of the displacement has been employed. We adopt the regularization method of \cite{tehrani2022bi} where the first- and second-order derivatives of displacement (strains and it's first-order derivative) are employed:  
	
	\begin{equation}
		\label{eq:loss_s}
		\begin{gathered}         
			L_S = L_{s1} + \gamma L_{s2}\\
			L_{s1} = |\varepsilon_{11}-<\varepsilon_{11}>|_1 + \beta|\varepsilon_{12}|_1+\frac{1}{2}|\varepsilon_{21}|_1+\frac{1}{2}\beta|\varepsilon_{22}|_1 \\	
			L_{s2} = \biggl \{|\ (\frac{\partial \varepsilon_{11}}{\partial a})|_1 + \beta |\ (\frac{\partial \varepsilon_{11}}{\partial l})\biggr.|_1 + \left. 0.5|\ (\frac{\partial \varepsilon_{22}}{\partial{a}})|_1+0.5\beta|\ (\frac{\partial \varepsilon_{22}}{\partial l}) |_1\right \}	
		\end{gathered}  
	\end{equation}
	where $L_{s1}$, $L_{s2}$ and $\gamma$ are the first-order smoothness loss, the second-order smoothness loss and the weight associated to the second-order smoothness loss, respectively. The $<.>$ denotes the mean operation, and the hyperparameters are tuned similar to \cite{tehrani2022bi}.

	We propose to add the PICTURE loss as the regularization constraint to improve the lateral displacement quality. Therefore, the final loss used to train the network is:
	\begin{equation}
		\label{eq:total_loss}
		Loss = L_D + \lambda_S L_S + \lambda_V L_V 
	\end{equation} 
	the hyperparameters $\lambda_S$ and $\lambda_V$, control the smoothness and PICTURE loss strength, respectively. It should be mentioned that the PICTURE regularization, similar to other forms of regularization, is only applied during the training and methods such as known operators \cite{maier2019learning} can be used to apply it during the test. 
	
	\subsection{Data Collection}
	\subsubsection{Experimental Phantom:}
	\label{exp_data}
	A tissue-mimicking breast phantom (Model 059, CIRS: Tissue Simulation \& Phantom Technology, Norfolk, VA) was employed for data collection. The elastic modulus of the phantom background was 20 kPa, and the phantom contained several inclusions having at least twice the elastic modulus of the background. An Alpinion E-Cube R12 research US machine (Bothell, WA, USA) with the sampling frequency of 40 MHz and the center frequency of 8 MHz was utilized for data collection of the training and test. To avoid data leakage, different parts of the phantom were imaged for training and test.
	\subsubsection{\textit{In vivo} data:} 
	A research Antares Siemens system by a VF 10-5 linear array was employed to collect data with the sampling frequency of 40 MHz and the center frequency of 6.67 MHz. Data was collected at Johns Hopkins Hospital from patients with liver cancer during open-surgical RF thermal ablation. The institutional review board approved the study with the consent of the patients. We selected 600 RF frame pairs of this dataset for the training of networks employed.

	\subsection{Network Architecture and Training Schedule}
	We employed MPWC-Net++ (publicly provided in \cite{tehrani2021mpwc}) which showed promising results on both optical flow and USE. This network was a modified variant of PWC-Net \cite{hur2019iterative} to address the high-frequency nature of RF data. The PWC-Net feature extraction part was modified to have less downsampling to reduce loss of information, and the cost volume search range was increased to address the low search range caused by reducing the number of downsampling in the feature extraction layers.   
	
	The publicly available pre-trained MPWC-Net++ weights were employed as the initial weights. The network was trained for 20 epochs with the learning rate of 20e-6, which was halved every five epochs, and an Nvidia A100 GPU with 40 Gb of memory was utilized for training the network. The tuned hyperparameters values can be found in the Supplementary Materials. The network weight and a demo code will be available online after the acceptance of the manuscript.
	
	\begin{figure}[t]
		\centering
		
		\begin{subfigure}{\textwidth}
			\centering
			\includegraphics[width=0.9\textwidth]{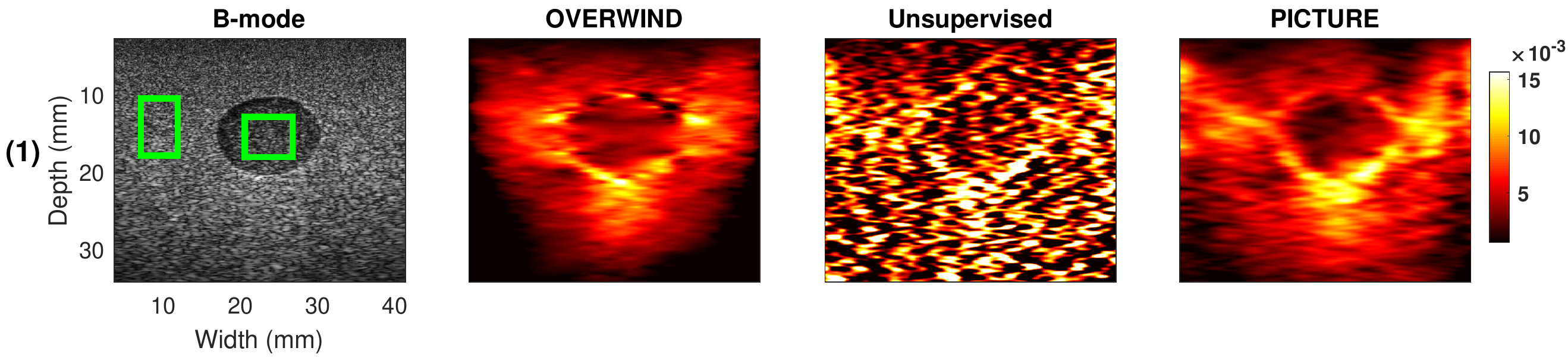}
			
		\end{subfigure}
		
		\begin{subfigure}{0.99\linewidth}
			\centering
			\includegraphics[width=0.9\textwidth]{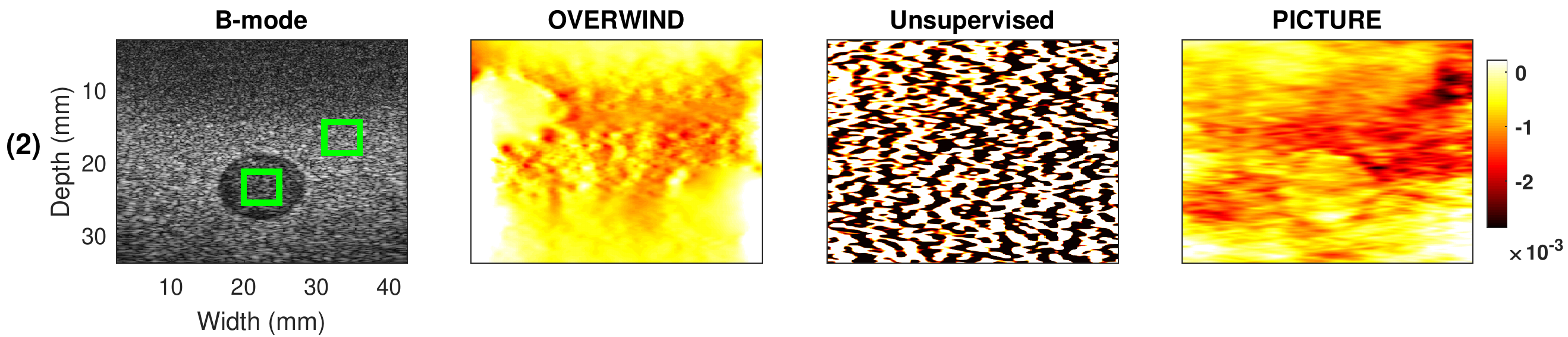}
			
		\end{subfigure}
		
		\begin{subfigure}{0.99\linewidth}
			\centering
			\includegraphics[width=0.9\textwidth]{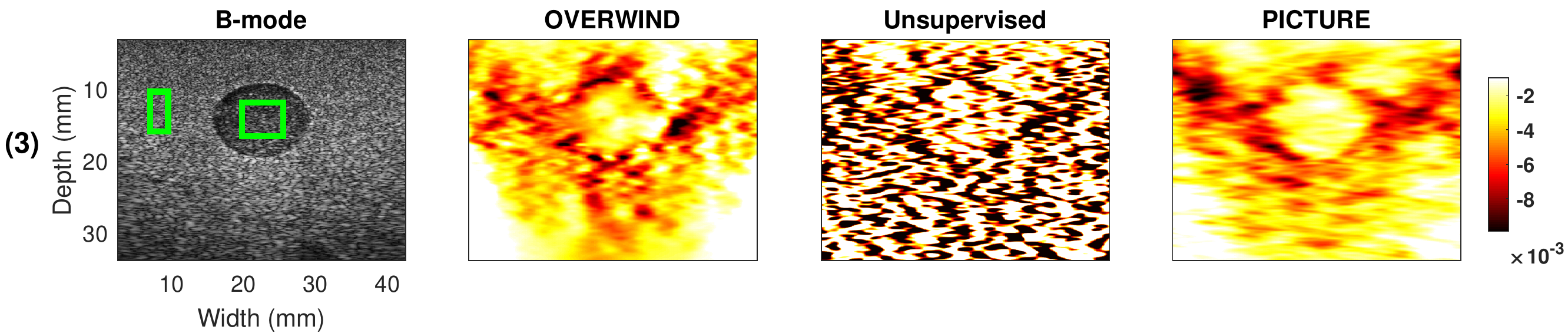}
		\end{subfigure}
		
		\begin{subfigure}{0.99\linewidth}
			\centering
			\includegraphics[width=0.9\textwidth]{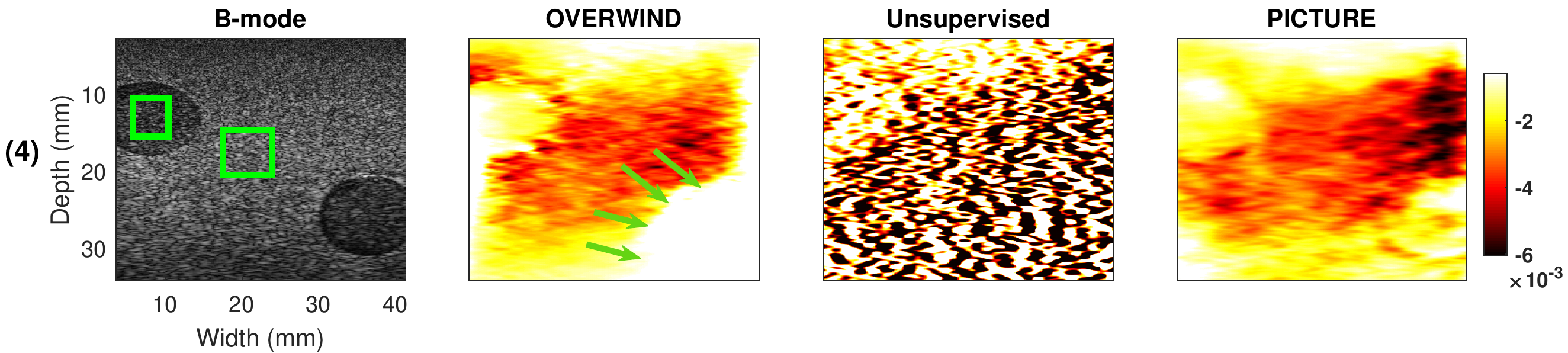}
			
		\end{subfigure}
		\caption{Lateral strain images in phantom experiments. Rows 1 to 4 correspond to different locations in the phantom.}
		\label{fig:phantom}
	\end{figure}
	\section{Results}
	PICTURE is compared to OVERWIND, an optimization-based method that uses the initial displacement obtained by Dynamic Programming \cite{mirzaei2019combining}. OVERWIND employs total variation for regularization and obtains high-quality sub-pixel displacement \cite{mirzaei2019combining}. In addition, PICTURE is compared to the regularized unsupervised training of MPWC-Net++ without PICTURE loss ($\lambda_V = 0$ in Eq \ref{eq:total_loss}) which the training loss function becomes similar to \cite{tehrani2020semi} but the difference is that a more recent network specifically designed for USE is employed.

	Contrast to Noise Ratio (CNR), and Strain Ratio (SR) are employed to quantitatively evaluate the compared methods. CNR and SR are defined as \cite{ophir1999elastography}:
	\begin{equation}
		\label{Eq:SRCNR}
		CNR = \sqrt{\frac{2(\overline{s}_{b}-\overline{s}_{t})^{2}}{{\sigma _{b}}^{2}+{\sigma _{t}}^{2}}},\quad \quad  SR =\frac{\overline{s}_{t}}{\overline{s}_{b}},
	\end{equation}
	where $\overline{s}_{X}$, and ${\sigma _{X}}^{2}$ are the mean and variance of strain in the target (subscript $t$) and background (subscript $b$) windows. Assuming that the target is stiffer than the background, lower SR, which is desired, represents a higher difference between the target and background. Also, CNR can provide a good intuition of the overall quality of strain images by combining both the mean and variance of the target and background windows. It should be noted that in order to improve the estimation of CNR and SR, they are computed for small overlapping patches inside the selected windows, and the mean and standard deviation are reported. 	
	\subsection{Experimental phantom results}
	We employ 2000 RF frame pairs of the dataset explained in Section \ref{exp_data}. Our main focus is on the lateral displacement ($\varepsilon_{22}$); therefore, we only present the lateral displacement results, and the axial ones are provided in the Supplementary Materials. The lateral strains of 4 experimental phantom image pairs are shown in Fig. \ref{fig:phantom}. It can be observed that the unsupervised method results in noisy strain images while, PICTURE provides lateral strain images of substantially higher quality by adding a regularization constraint to the unsupervised loss function. OVERWIND performs better than the unsupervised method but estimates inaccurate lateral strain. For instance, in phantom result (2), the inclusion is not detectable by the strain image obtained by OVERWIND whereas, it can be identified by the strain image obtained by PICTURE. Furthermore, in sample (4), the lateral strain of OVERWIND in the highlighted region is incorrect since it has a positive value. However, the correct lateral strain is negative (the histogram of EPR values of compared methods are given in the Supplementary Materials). 
	It should be noted that lateral strain estimation is a more challenging task than the axial one. The evaluated methods provide high-quality axial strain images given in the Supplementary Materials.

	\subsection{\textit{In vivo} results}
\begin{figure}	
		\begin{subfigure}{\textwidth}
			\centering
			\includegraphics[width=0.85\textwidth]{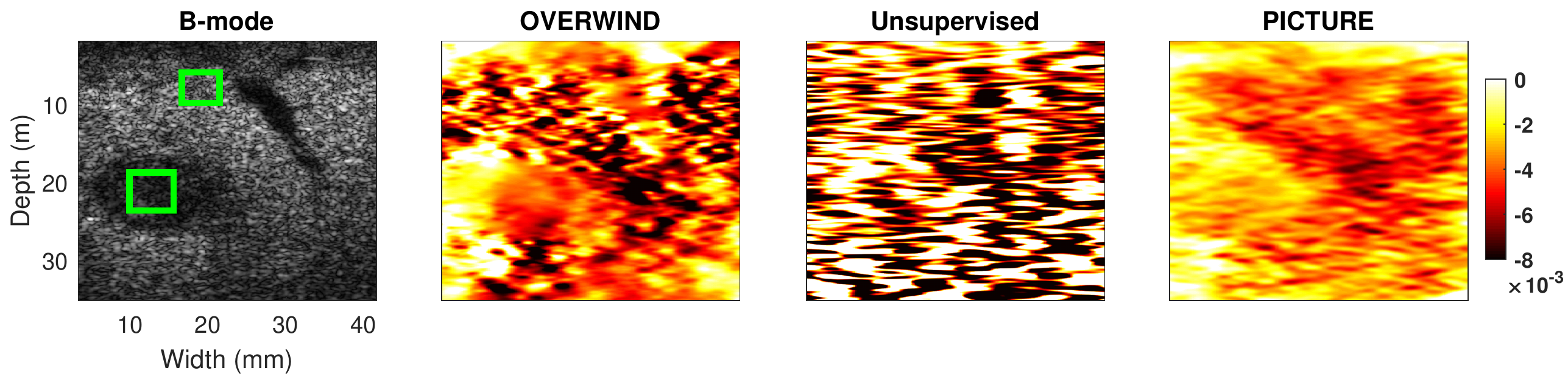}
			
		\end{subfigure}
		
		\begin{subfigure}{\linewidth}
			\centering
			\centering
			\includegraphics[width=0.6375\textwidth]{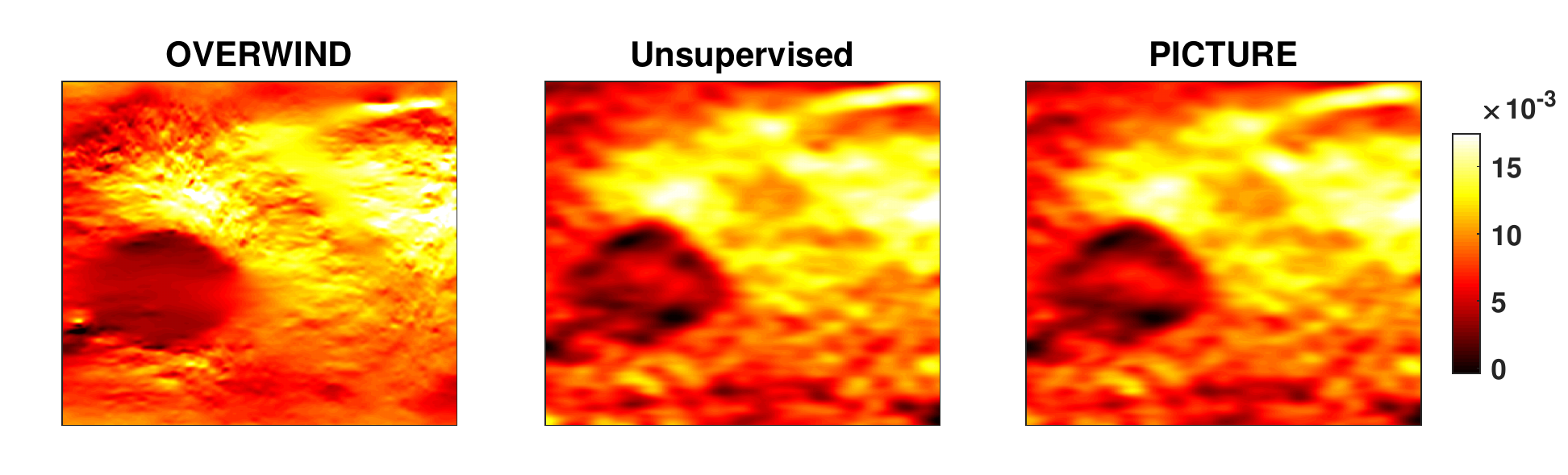}
		\end{subfigure}
		\caption{B-mode and strain images of a patient with liver cancer. Lateral and axial strain images are shown in the top and bottom rows, respectively.}
		\label{fig:invivo}
	\end{figure}

	The network was trained with (PICTURE) and without (Unsupervised) the proposed regularizer using \textit{in vivo} dataset. The lateral strain images are illustrated in Fig. \ref{fig:invivo} (top), and the axial strain images are given as a reference (bottom) for one patient data before liver ablation. It can be observed that PICTURE substantially improves the lateral strain images of the unsupervised method while their axial strain images are virtually similar.  
	\subsection{Quantitative results}
	The CNR and SR of the experimental phantom and \textit{in vivo} data results are given in Table \ref{tab:my-table}. In terms of CNR metric, PICTURE substantially outperforms OVERWIND and the unsupervised method. For \textit{in vivo} data, PICTURE increases the CNR of the unsupervised method from the mean value of $0.73$ to $4.27$ and has substantially higher CNR than OVERWIND (1.12). In terms of SR, PICTURE has the lowest SR in 3 cases, and the unsupervised method has better SR in 2 cases (phantom results 3 and 4). However, the visual results and CNR values indicate that the unsupervised method does not have high-quality lateral strains in those cases. For \textit{in vivo} data, OVERWIND has SR value of $1.06$ which means that it incorrectly estimates the target (tumor) has a higher strain value than the background. Whereas unsupervised and PICTURE can detect the lower strain value of the tumor (SR<1). Also, PICTURE has a lower standard deviation ($0.06$) compared to unsupervised method ($0.42$). 
	
	\begin{table}[t]
		\resizebox{0.99\textwidth}{!}{
			\caption{Mean and standard deviation of CNR and SR of experimental phantom results. OV, UN, and PIC represent OVERWIND, the unsupervised method, and PICTURE, respectively. Statistical significance is achieved using Friedman's test (\textit{p}-value$<0.01$).}
			\label{tab:my-table}
			\begin{tabular}{@{}c|cccccc|cccccc@{}}
				\toprule
				& \multicolumn{6}{c|}{CNR (higher is better)}              & \multicolumn{6}{c}{SR (lower is better)}               \\ \midrule
				Phantom& OV  && UN && PIC    && OV  && UN && PIC   \\\midrule
				
				(1) &3.16$\pm$1.45 && 0.80$\pm$0.40    && \textbf{6.22$\pm$1.45}  && 0.68$\pm$0.09 && 0.57$\pm$0.17    && \textbf{0.35$\pm$0.07} \\
				\rowcolor[HTML]{C0C0C0} 
				(2) & 1.21$\pm$0.73 && 0.38$\pm$0.30    && \textbf{4.72$\pm$0.50}  && 0.86$\pm$0.12 && 10.00$\pm$6.80   && \textbf{0.50$\pm$0.07} \\
				
				(3) & 4.19$\pm$0.83 && 0.84$\pm$0.38    && \textbf{5.80$\pm$1.46}  && 0.42$\pm$0.07 && \textbf{0.34$\pm$0.18}    && 0.36$\pm$0.05 \\
				\rowcolor[HTML]{C0C0C0}
				(4) & 9.38$\pm$2.53 && 1.44$\pm$0.32    && \textbf{12.45$\pm$0.38} && 0.35$\pm$0.06 && \textbf{0.21$\pm$0.12}    && 0.38$\pm$0.1  \\\midrule 
				\textit{in vivo} data&1.12$\pm$0.84&&0.73$\pm$0.54&&\textbf{4.27$\pm$1.12}&&1.06$\pm$0.30&&0.83$\pm$0.42&&\textbf{0.66$\pm$0.06}\\\bottomrule
		\end{tabular}}
	\end{table}
	\section{Conclusions}
	Advances in several fields such as inverse reconstruction of elasticity modulus and imaging of the effective Poisson's ratio have been hindered because of the low quality of the lateral displacement estimation in USE. This paper takes a step to resolve those issues by incorporating governing laws of physics in USE to improve the lateral strain. 
	We employed the axial displacement to improve the lateral one by considering the constraint imposed by the feasible range of EPR. A new regularization term called PICTURE loss was added to the loss function of unsupervised training. Experimental phantoms and \textit{in vivo} data were employed to validate the proposed method. The visual and quantitative evaluations confirmed the effectiveness of PICTURE in improving the lateral strain quality.  
	\section{Acknowledgment}
	The authors would like to thank Drs. E. Boctor, M. Choti and G. Hager for providing us with the \textit{in vivo} patients data from Johns Hopkins Hospital.

	\bibliographystyle{splncs04}
	\bibliography{IEEEfull}
	
\end{document}


\title{Supplementary Materials for Physically Inspired Constraint for Unsupervised Regularized Ultrasound Elastography\thanks{Supported by Natural Sciences and Engineering Research Council of Canada (NSERC) Discovery Grant. The Alpinion ultrasound machine was partly funded by Dr. Louis G. Johnson Foundation.}}
	%
	%
		\author{Ali K. Z. Tehrani$^{1,2}$, and Hassan Rivaz$^{1,3}$}
	\authorrunning{Ali K. Z. Tehrani}
	%
	\institute{Department of Electrical and Computer Engineering, Concordia University, Canada\and 
		\email{a_kafaei@encs.concordia.ca}  \email{ $^3$ hrivaz@ece.concordia.ca}}
	\maketitle              

	\begin{table}[t]
		\caption{Hyperparameters values and descriptions. The numbers inside the parenthesis indicate the values used for \textit{in vivo} data. }
		\label{tab:my-table}
		\begin{tabular}{ccc}
			\hline
			Hyper-Parameter & Description                                                                  & Value      \\ \hline
			$v_{emin}$      & \begin{tabular}[c]{@{}c@{}}minimum range of \\ accepted EPR\end{tabular}     & 0.1        \\
			\rowcolor[HTML]{C0C0C0} 
			$v_{emax}$      & \begin{tabular}[c]{@{}c@{}}maximum range of \\ accepted EPR\end{tabular}     & 0.6        \\
			$\beta$      & \begin{tabular}[c]{@{}c@{}}related to ratio of\\ axial to lateral distance\end{tabular}                     & 0.1     \\
			\rowcolor[HTML]{C0C0C0} 
			$\gamma$        & \begin{tabular}[c]{@{}c@{}}weight of second-order \\ smoothness\end{tabular} & 8          \\
			$\lambda_{vs}$  & smoothness weight of EPR                                                     & 0.25       \\
			\rowcolor[HTML]{C0C0C0} 
			$\lambda_S$  & \begin{tabular}[c]{@{}c@{}}weight of smoothness \\ regularizer\end{tabular}                                 & 2.2 (3) \\
			$\lambda_V$     & weight of PICTURE loss                                                       & 0.2 (0.05) \\
			\rowcolor[HTML]{C0C0C0} 
			Not assigned & \begin{tabular}[c]{@{}c@{}}patch axial size inside \\ the windows for \\ CNR, SR calculation\end{tabular}   & 100     \\
			Not assigned & \begin{tabular}[c]{@{}c@{}}patch lateral size inside \\ the windows for \\ CNR, SR calculation\end{tabular} & 30      \\ \hline
		\end{tabular}
	\end{table}	
	
	\begin{figure}[h]
		\includegraphics[width=0.55\textwidth]{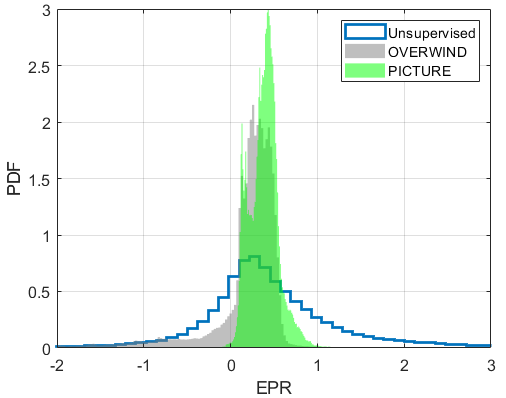}
		\caption{Histogram of EPR values of phantom result 4. Although, PICTURE limits the range of Poisson’s ratio, there is no guarantee it is exactly in the considered range during the test time.}
	\end{figure}	
	
	\begin{figure}[h]

		\begin{subfigure}{\textwidth}
			\hspace{1.2cm}
			\includegraphics[width=0.75\textwidth]{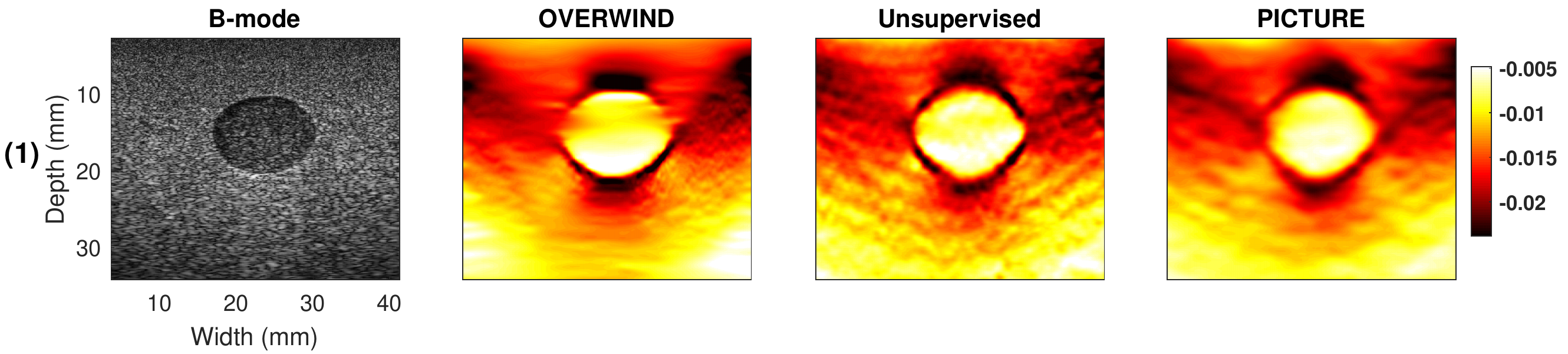}\hspace*{\fill}
			
		\end{subfigure}
		
		\begin{subfigure}{0.99\linewidth}
			\hspace{1.2cm}
			\includegraphics[width=0.75\textwidth]{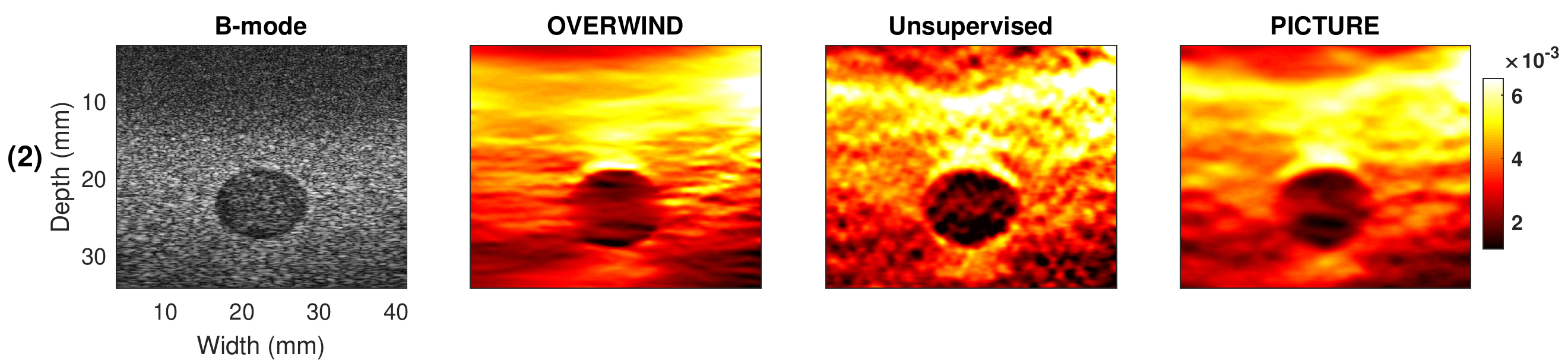}\hspace*{\fill}
			
		\end{subfigure}
		
		\begin{subfigure}{0.99\linewidth}
			\hspace{1.2cm}
			\includegraphics[width=0.75\textwidth]{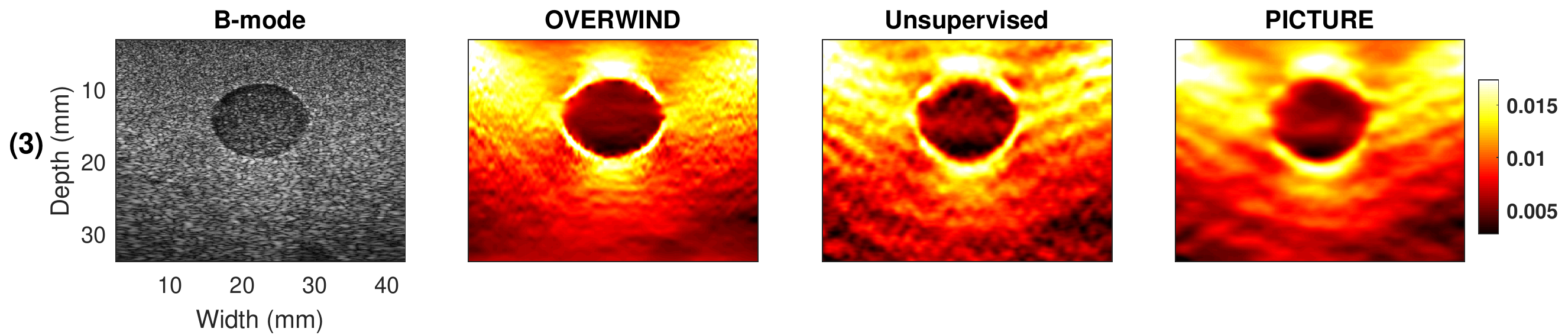}\hspace*{\fill}
		\end{subfigure}
		
		\begin{subfigure}{0.99\linewidth}
			\hspace{1.2cm}
			\includegraphics[width=0.75\textwidth]{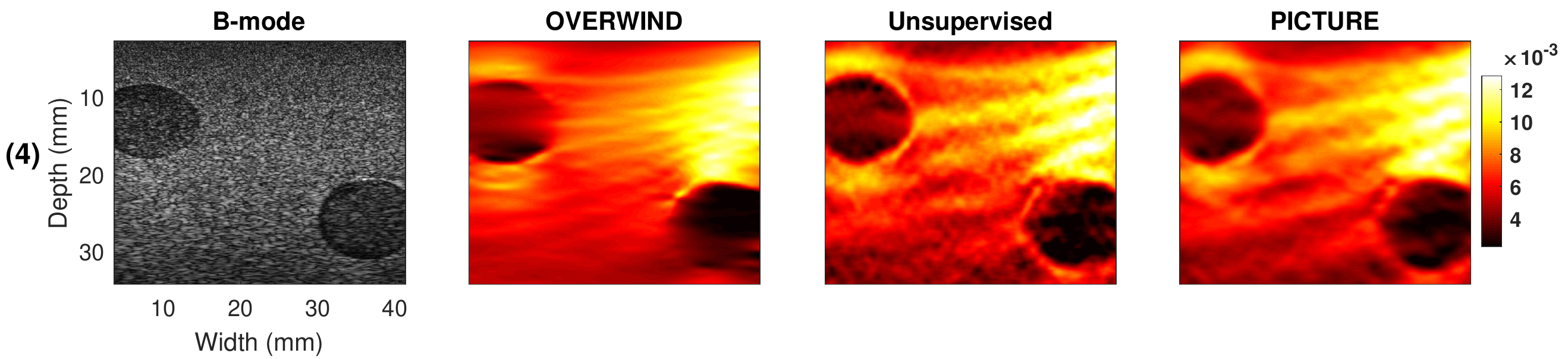}\hspace*{\fill}
			
		\end{subfigure}
		\vspace{-0.25cm}
		\caption{Experimental phantom axial strain results.}
		\label{fig:phantom}
	\end{figure}